\documentstyle[12pt,fleqn,epsf]{ioplppt} 

\begin{document}

\title{Non-Singular Cosmological Models in String Gravity
       with Second Order Curvature Corrections}

\author{S.O. Alexeyev\ftnote{1}{alexeyev@grg2.phys.msu.su},
        A.V. Toporensky\ftnote{2}{lesha@sai.msu.su},
        V.O. Ustiansky\ftnote{3}{ustiansk@sai.msu.su}}

\address{Sternberg Astronomical Institute,
         Moscow State University, \\
         Universitetskii Prospect, 13,
         Moscow 119899, Russia}

\begin{abstract}
    We investigate FRW  cosmological  solutions in the theory of
modulus  field  coupled to gravity through a Gauss-Bonnet  term.
The explicit  analytical  forms  of  nonsingular asymptotics are
presented for power-law and exponentially steep modulus coupling
functions. We study  the influence of modulus field potential on
these asymptotical regimes  and find some forms of the potential
which do not destroy the nonsingular behavior. In particular, we
obtain that  exponentially steep coupling functions arising from
the  string  theory  do  not allow nonsingular  past  asymptotic
unless modulus field potential  tends  to zero for modulus field
$\psi \to \pm  \infty$. Finally, the modification of the chaotic
dynamics  in the  closed  FRW universe due  to  presence of  the
Gauss-Bonnet term is discussed.
\end{abstract}


\section{Introduction}

    A great progress in constructing  a  theory  of all physical
interactions occurred over the past few  decades. Standard model
of the  particle  physics  unified  electro-magnetic,  weak  and
strong  interactions  into  one  theory at the energy  scale  of
$10^{16}$ GeV. Unfortunately, methods of Quantum Field Theory do
not result with a huge progress when applied to the quantization
of General Relativity. Therefore to avoid  problems in cosmology
and  black  hole  physics  a  new  approach (String Theory)  was
constructed to study the topological structure of the space-time
in the near Planckian regions \cite{s2,s1}.

    In  the  perturbational approach developed at a first  stage
string theory predicts  the Einstein equations to be modified by
the  higher  order curvature  corrections  in  the  range  where
curvature of space-time has near-Planckian values (the situation
which  took  place  at   very   early  stages  of  our  Universe
evolution). At  the present time  the form of these higher order
curvature corrections  in  the  string  effective  action is not
investigated completely. We do not know the general structure of
the  expansion  and, hence, the direct summation is  impossible.
But in the perturbative expansion the  most important correction
is  the  second order curvature one,  it  is the product of  the
Gauss-Bonnet  with   dilatonic   and   moduli   terms.  In  four
dimensional  space-time  Gauss-Bonnet term  represents  a  total
divergence but its  combination  with a dilatonic (moduli) field
makes  the  contribution of the second order  correction  to  be
dynamical. The investigations in the  frames  of  the  discussed
model were performed formerly and a set of  new cosmological and
black hole  type  solutions  was found \cite{m1,k1,m3,sts} which
provide  new  types of space-time topology \cite{m2}. All  these
effects appear in  the higher order curvature string gravity and
are absent in the minimal Einstein gravity.

    At the  next stage of the  string theories evolution  it was
proved that all five independent string  theories were connected
by the different dual transitions  and  they all were a part  of
M-theory \cite{x1}, which in low energy limit gives rise to 11th
dimensional supergravity.  Using  duality transitions one has an
opportunity to work with a  non-perturbative  theory  but at the
current  moment   non-perturbative   M-theory  is  only  at  the
beginning  of  its  way,  its study methods are  not  completely
created \cite{tj1}.

    So, in order  to  make a  little  step in understanding  the
space-time  structure,   it   is   possible   to   work  in  the
perturbational approach and try to study new topological effects
of  low  energy string action with the  higher  order  curvature
corrections.  Here  it   is  necessary  to  note  that  all  the
conclusions  obtained  in the  near  Planckian  region  must  be
treated  as  only preliminary directions on some effects,  later
they have to be verified  by  quantum  gravity calculations. So,
the most convenient form of the string gravity action is (we use
the units where $m_{PL}/\sqrt{8 \pi}=1$):
\begin{eqnarray}\label{a0}
S = \frac12 \int d^4 x \sqrt{-g} \biggl[ - R +
\partial_\mu \psi \partial^\mu \psi + \partial_\mu
\phi \partial^\mu \phi
+ \lambda e^{-2 \phi} S_{GB} + \delta \xi(\psi)
S_{GB}  \biggr] ,
\end{eqnarray}
    where $R$ is Ricci  scalar,  $\psi$ is modulus field, $\phi$
is dilaton,  $\lambda$  ($\delta)$ is dilatonic (modulus) string
coupling constant. Their values depend upon the concrete type of
the string  theory.  Second order curvature correction $(\lambda
e^{-2\phi} + \delta \xi)  S_{GB}$  represents the product of the
coupling   functions   $e^{-2   \phi}$   and   $\xi(\psi)$   and
Gauss-Bonnet combination $S_{GB} = R_{ijkl} R^{ijkl}  - 4 R_{ij}
R^{ij} + R^2$.

    The importance of moduli field in cosmology is a consequence
of the  fact that the  sing of  $\delta$ is not  fixed from  the
first  principles  and  can  be  either   positive  or  negative
depending upon the  detail  structure of the theory. Antoniadis,
Rizos and  Tamvakis  \cite{Antoniadis:1994jc} have shown that in
the  case  of  negative  $\delta$  a  new class of  cosmological
solutions appeared with  nonsingular  past asymptotic and smooth
transition  to  ordinary Friedmann Universe. In the next  papers
\cite{m3,Rizos:1994rt,Kanti:1998jd} such  kind of solutions  was
constructed for metrics with zero or positive spatial curvature.
We recall that in Einstein  theory  with a scalar field all  the
possible past asymptotics for flat models  are singular (thought
if scalar field potential is very steep, trajectory can approach
a  singularity  in  a  complicated way, see  \cite{Foster}).  In
closed models  there  are nonsingular periodical and aperiodical
solutions but their measure in  the  initial  condition space is
equal to zero\cite{Corn-Shel}.  The  presence of modulus term in
action (\ref{a0}) leads to appearance  of  a  set of nonsingular
solutions  with  finite  (and  sufficiently  large)  measure  of
required initial conditions.

    The  presence   of   dilatonic   field   does   not   change
qualitatively these  non-singular solutions. The authors of Ref.
\cite{Rizos:1994rt}  have  proposed  to  neglect  the  dilatonic
contribution at all to simplify the analysis and, hence, to show
the phenomenon  of  nonsingular cosmological behavior induced by
modulus field in its ``pure'' form. We follow  this proposal and
consider only modulus-dependent part in addition to the Einstein
term in (\ref{a0}). $\xi(\psi)$  in  string theory is defined in
terms of Dedekind  function and is exponentially steep for $\psi
\to \pm  \infty$ \cite{m3}. We do  not restrict ourself  by this
particular  case  and  analyze  a much wider class  of  coupling
functions.

    The concrete form  of  nonsingular asymptotic depends on the
function $\xi(\psi)$. In \cite{Rizos:1994rt} the particular case
$\xi \sim \psi^2$ is investigated both  by  the  analytical  and
numerical    methods.    The   asymptote    for    the    closed
Friedmann-Robertson-Walker (FRW) metric has the form
\begin{eqnarray}
a = const = \sqrt{-12 \delta} \;,\qquad\qquad
\psi = \psi_0+\sqrt{-\frac{1}{2 \delta}}t \;,
\end{eqnarray}
    where $a$ is the scale factor of the FRW metric.

    One  can  see  that  the  nonsingular  dynamics  on  $a$  is
accompanied with the growth of the absolute value of the modulus
field.  The same  feature  presents in the  FRW  flat case  (see
below). It  was  noticed  in  \cite{Rizos:1994rt},  that in such
situation the modulus field potential, omitted in (\ref{a0}) can
be important.  As  nonzero  modulus  field  potential appears in
theories  with  supersymmetry  breaking \cite{Tseytlin} and  the
scale of SUSY  breaking is very  small in the  Planckian  scale,
therefore, modulus field is very light. But the unbounded growth
of modulus field causes the necessity to take  into account this
potential  term  in  asymptotes  (2), and such  puzzle  requires
special additional analysis. So, in this paper we consider
the action in the form:
\begin{eqnarray}\label{a1}
S = \frac12 \int d^4 x \sqrt{-g} \biggl[ - R +
\partial_\mu \psi \partial^\mu \psi
+ \delta \xi(\psi) S_{GB} + 2\ V(\psi) \biggr] ,
\end{eqnarray}
where $V(\psi)$ is the modulus field potential.

    The structure of  our paper is  the following: in  Sec.2  we
present nonsingular asymptotics in FRW  flat  case  for  several
types of  modulus  coupling function $\xi(\psi)$ and investigate
the  influence  of  the  modulus  field  potential upon them.  A
particular attention is  paid  to possibility of the nonsingular
regime to  survive in the  presence of the nonzero potential. In
Sec.3 this analysis is extended in a more qualitative way to the
positive curvature FRW metric.  Sec.4  provides a summary of the
results obtained.

\section{Flat case}

    In this section we consider dynamics of a  flat FRW universe
with the metric
\begin{eqnarray}
d s^2 = dt^2 - a^2(t) \biggl( dx^2 + dy^2 + dz^2 \biggr)\; .
\end{eqnarray}

    After introducing a Hubble parameter $h=\dot{a}/a$ the field
equations are
\begin{eqnarray}
& &3 h^2 \biggl(1+4\delta\dot\xi h\biggr)-\dot{\psi}^2/2- V=0\;,
\label{constr}\\
& &2\biggl(\dot h+h^2\biggr)\biggl(1+4\delta\dot\xi h\biggr)+
h^2\biggl(1+4\delta\ddot\xi\biggr)+\dot\psi^2/2- V=0\;,
\label{eqmot1}\\
& &\ddot\psi+3h\dot\psi+ V' -12\delta\xi'h^2\biggl(\dot
h+h^2\biggr)=0\;.
\label{eqmot2}
\end{eqnarray}
    The prime  (dot)  denotes  differentiation  with  respect to
$\psi$ ($t$).

    In this section our aim is to construct the explicit form of
nonsingular    asymptotic    solutions    of    the    equations
(\ref{constr})--(\ref{eqmot2}). In  order  to attain this aim we
consider the contraction of the model,  therefore, the expansion
of the flat  universe  corresponds to  the  time reverse of  the
solutions listed below.

\subsection{The case $V(\psi)=0$}

    Previously      Antoniadis,      Rizos     and      Tamvakis
\cite{Antoniadis:1994jc,Rizos:1994rt}     have     shown    that
nonsingular solutions can exist only for $\delta<0$, which would
be assumed from now on.

    In our investigation  we  assume a modulus coupling function
to be  represented by a power law ($\xi(\psi)=\psi^{n+1}/(n+1)$)
or  asymptotically  exponential   ($\xi(\psi)\sim   e^{|\psi|}$)
forms. We do  not  consider functions $\xi(\psi)$ growing slower
than $\psi^2$  because  for $|\psi|\to\infty$ these functions do
not  provide  the  violation  of  both  strong  and  weak energy
conditions \cite{Antoniadis:1994jc,Kanti:1998jd}.

    To construct  nonsingular  asymptotics  it  is  possible  to
neglect the  first  order  curvature  term  in (\ref{constr}) in
comparison  with  the  second  one. So, the  reduced  constraint
equation is
\[
\frac{\dot \psi}{\xi'} = 24\delta h^3\;.
\]

    The left hand  side (LHS) of the reduced constraint equation
can be  easily  integrated for considered functions $\xi(\psi)$.
The result is
\vskip4mm
\begin{tabular}{rl}
LHS $\sim$ &
$\left\{ \begin{tabular}{ll}
$\ln \psi\;$ ,& $\quad\xi(\psi)\sim\psi^2$\;;\\
$\psi^{1-n}\;$, & $\quad\xi(\psi)\sim\psi^{n}\;,\;\;n>2$\;;\\
$e^{-|\psi|}$\;, & $\quad\xi(\psi)\sim e^{|\psi|}$\;;
\end{tabular}
\right.$
\end{tabular}
\vskip4mm
Thus, these cases must be treated separately.

    For   $\xi(\psi)=\psi^2/2$   anticipating   the   asymptotic
solution
\[
h=At^{\alpha}\;,\qquad \psi=Be^{\beta t}
\]
    and  substituting   this   ansatz  into  the  reduced  field
equations one obtains
\begin{eqnarray}
h(t) &\sim& -\sqrt{\frac{-5}{48\delta}}\; , \nonumber \\
\psi(t) &\sim& A\exp\left(24\delta h^3 t\right)\; , \quad A=const.
\label{psi2}
\end{eqnarray}
    The scale factor  $a(t)$ in this case exponentially tends to
zero.

    For  $\xi(\psi)=\psi^{(n+1)}/(n+1)$  anticipation   of   the
asymptotical solution in a form
\[
h=At^{\alpha}\;,\qquad \psi=Bt^{\beta}
\]
    leads to two different cases:  $1<n<5$  and  $n\ge 5$\@. For
the first one the explicit form of the solution is:
\begin{eqnarray}
h(t) &\sim& -\frac{n-5}{5(n-1)}t^{-1}\; ,  \nonumber \\
\psi(t) &\sim& \pm\left(\frac{125(n-1)^2}{12\delta(n-5)^3}\right)^
\frac{1}{n-1} t^\frac{2}{n-1}\;,
\label{nl5}
\end{eqnarray}
    the scale  factor on this solution  tends to zero  vie power
law $a(t)\sim  t^{-\frac{n-5}{5(n-1)}}$\@. For the second one we
have:
\begin{eqnarray}
h(t) &\sim& A t^{-\frac{2n}{n+5}}\; , \quad A=const<0\; ,
\nonumber \\
\psi(t) &\sim& \pm\left(\frac{5}{24\delta(n+5)A^3}
\right)^{\frac{1}{n-1}}
t^{\frac{5}{n+5}}\; .
\label{ng5}
\end{eqnarray}
    For $n>5$ the  scale factor $a(t)$ asymptotically tends to a
constant positive value.

    The case $\xi(\psi)\sim e^{|\psi|}$, which is the asymptotic
form of function $\xi(\psi)$  arising  in string theory has been
previously investigated  in detail in  \cite{Antoniadis:1994jc}.
In  the  cited paper three possible asymptotical solutions  have
been listed but only one of them corresponded to the contraction
of  the  Universe. In  order  to make  a  complete reference  of
nonsingular asymptotics we specify it in our notations here:
\begin{eqnarray} h(t)  &\sim&  A  t^{-2}\;  ,
\quad  A=const<0\;  ,  \nonumber  \\ |\psi(t)| &\sim& 5\ln  t  +
\ln\frac{5}{24\delta A^3} \; .
\label{ke}
\end{eqnarray}

\subsection{The case $V(\psi)\ne0$}

    Here  it  worth  to  notice  that  for  arbitrary  functions
$\xi(\psi)$ and $V(\psi)$ the action (\ref{a1}) allows rescaling
with respect to $\delta$:
\begin{equation}
t\to \sqrt{|\delta|}t\;,\qquad h\to
\frac{h}{\sqrt{|\delta|}}\;, \qquad
V\to \frac{V}{\sqrt{|\delta|}}\;.
\label{scaling}
\end{equation}
    Hence,  it  is possible to eliminate the parameter  $\delta$
from    the    equations    (\ref{constr})--(\ref{eqmot2})    by
substituting  $\delta\equiv  -1$ which will be assumed from  now
on.

    First of all it is necessary to check whether it is possible
to introduce the  potential $V(\psi)$ which does not violate the
asymptotical  solutions  obtained in  the  previous  subsection.
Substituting  these   asymptotical  solutions  into   constraint
equation  (\ref{constr})  one  can  see  that  all the items  in
constraint    tend    to    zero   with   $t\to\infty$    unless
$\xi(\psi)=\psi^{n+1}/(n+1)$ for $n\le 3$.

    For $\xi(\psi)=\frac{\psi^2}{2}$ the  asymptotical  solution
in   the   form   (\ref{psi2})   provides  all  the   terms   in
(\ref{constr}) to be of second power with respect to $\psi$. So,
the nonsingular  solutions  are  not  violated  if the potential
grows  slowly   than  $\psi^2$  at  $|\psi|\to\infty$.  For  the
particular potential $V(\psi)=m^2 \psi^2/2$ a generalization  of
the   solution   (\ref{psi2})  ($h=const\;$,   $\psi=Ce^{At}\;$,
$A=const\;$,  $C=const$)  can  be  substituted  into  the  field
equations which leads to
\begin{eqnarray}
A^2+24 h^3 A+m^2&=&0\;,\nonumber\\
A^2+3 h A +12h^4+m^2&=&0\;.\nonumber
\end{eqnarray}
    This  system  has real solutions only for  $m  <  m_0\approx
0.32$ which implies that the nonsingular asymptotes are violated
unless $m<m_0$.  The  potentials  steeper  than  $\psi^2$ always
violate the nonsingular asymptotic under consideration.

    For $\xi(\psi)=\frac{\psi^4}{4}$ the  asymptotical  solution
(\ref{nl5}) gives $\dot \psi  =  const$\@. So, the only possible
type  of  potential which  does  not  violate  this  nonsingular
asymptotic  is   the   asymptotically  flat  one  ($V(\psi)  \to
\Lambda=const$ for $|\psi| \to \infty$). Substituting the ansatz
$h(t)\sim Ct^{-1}$, $\psi(t)\sim At$ ($C=const$, $A=const$) into
the      field     equations      the      constraint      gives
$C^3=-\frac{A^2+\Lambda}{24  A^4}$\@. Being  combined  with  the
equation (\ref{eqmot2}) it results
\[
 \left(2\Lambda-5
A^2\right)^3+24 A^4 \left(A^2+2\Lambda\right)^2=0\;,
\]
    which has real roots only  if  $\Lambda  <  \Lambda_0\approx
0.42$\@. This means  that  the solutions remains nonsingular for
$\Lambda < \Lambda_0$.

    As it  is stated above, for $\xi(\psi)=\psi^{n+1}/(n+1)$, $n
>  3$  and  for  exponentially  steep  $\xi(\psi)$ all terms  in
(\ref{constr}) tend  to  zero  which  implies  that an arbitrary
asymptotically   nonvanishing   modulus    potential    destroys
corresponding solutions (\ref{ng5}),(\ref{ke}).

    At the end of this section we describe singular solutions of
our system. Previously it was shown \cite{Rizos:1994rt} that for
$V(\psi)=0$ there was  only  one singular asymptotic. It belongs
to the class  $|h|\to \infty$, $\dot\psi$ --- finite of singular
solutions.  It  turns out that the introduction  of  a  positive
potential  $V(\psi)$   which  remains  finite  with  it's  first
derivative  for  finite  $\psi$  will not lead to  new  singular
asymptotics of  this type. To  prove this we apply the procedure
described in the above cited paper.

    The     equations     of     motion    for    the     system
(\ref{constr})-(\ref{eqmot1}) may be rewritten as:
\begin{eqnarray}
\dot\psi&=&12\delta\xi'h^3\pm\sqrt{144\delta^2\xi'^2h^6+6h^2-2V}\;,
\label{dpsi}\\
h'&\equiv&\frac{dh}{d\psi}=-\frac1{\dot\psi}\frac AB\;,
\label{dh}
\end{eqnarray}
where
\begin{eqnarray}
A&=&48\delta\xi''\dot\psi^4h^4+5\dot\psi^4h^2+36h^6+36\dot\psi^2
h^4 +h^2V^2
+12h^3\dot\psi V'-2\dot\psi^3 h V'\nonumber\\
& &-2V\left(12h^4+6\dot\psi^2h^2+2h\dot\psi V'\right)\;,
\label{AA}\\
B&=&36h^4-12h^2\dot\psi^2+5\dot\psi^4+12V\left(
\dot\psi-2h^2\right)+4V^2\;.
\label{BB}
\end{eqnarray}
    The  equation  (\ref{eqmot2})  can  be  excluded  since  the
equations (\ref{constr})--(\ref{eqmot2}) are not independent.

    For the  singular  solution  under consideration, keeping in
mind finiteness  of  $\psi$  and  aforementioned restrictions on
$V(\psi)$,  we  can  neglect  the  term  $V(\psi)$  in  equation
(\ref{dpsi}) and,  therefore,  we  only  need  to study equation
(\ref{dh}).  There  are three  possibilities: $\delta\xi'h^2\sim
O(1)$, $\delta\xi'h^2\ll O(1)$ and  $\delta\xi'h^2\gg  O(1)$. In
the first case we  obtain  from (\ref{dpsi}) that $\dot\psi \sim
\alpha h$\@, $\alpha=const$\@. Substituting  this  solution into
equation  (\ref{AA}) we  can  see that it  is  not necessary  to
proceed the separate cases for $\xi''$\@. It is enough to notice
that  we  can neglect  all  $V$-terms (the  eldest  of them  are
proportional  to  $h^4$)  with  respect  to  terms  proportional
$h^6$\@.  In  equation  (\ref{BB})  we  also   can  neglect  all
$V$-terms (the eldest of them  are  proportional  to $h^2$) with
respect to terms proportional to $h^4$\@. Thus, this case can be
treated  as  with  $V=0$.  In  the   second   case   we   obtain
$\dot\psi\sim\pm\sqrt{6}h$ and we treat it in a complete analogy
with the previous one. For $\delta\xi'h^2\gg O(1)$ there are two
possibilities:   $\dot\psi_+\sim\xi'h^3$   and   $\dot\psi_-\sim
\frac{1}{\xi'h}$\@.   For   $\dot\psi=\dot\psi_+$   the   eldest
$V$-term in  equation  (\ref{AA})  is  proportional  to $h^{10}$
which  can  be  neglected  with respect to $h^{14}$  terms.  For
$\dot\psi=\dot\psi_-$  the  eldest  $V$-term  in  (\ref{AA})  is
proportional to $h^4$ which can be  neglected  with  respect  to
$h^6$ term.

    So, the asymptotic \cite{Rizos:1994rt}
\begin{eqnarray}
h(t) &\sim& \left(t-t_0\right)^{-1} \to -\infty\; ,
\quad t\to t_0-0\; ,
\quad t_0=const\; , \nonumber\\
\label{mcomm}
\psi(t) &\sim& \psi_0-\frac{1}{8\delta\xi'}\left(t-t_0\right)^2\; ,
\quad
\psi_0=const\;
\end{eqnarray}
    does not  depend  on  $V(\psi)$  if  $V(\psi)$ satisfies the
aforementioned requirements.

    If $V=0$ the signs of $h$ and $\dot\psi$ are crucial as they
are conserved independently. For $V\ne 0$ the sign of $\dot\psi$
is not conserved. So, solution can be singular irrespectively of
the initial sign of $\dot\psi$ in contrast to the case $V=0$.

\section{The positive curvature case}

    The FRW metric for the positive spatial curvature case is
\[
ds^2=dt^2-a^2(t)d\Omega^2\;,
\]
    where $d\Omega^2$ is the volume element of the 3-dimensional
sphere, field equations are
\begin{equation}
2 \frac{\ddot a}{a}(1+4 \delta \dot \xi \frac{\dot a}{a})+
\left( \frac{\dot a^2}{a^2}+ \frac{1}{a^2}\right)(1+ 4
\delta \ddot \xi)=
V(\psi)-\frac{\dot \psi^2}{2}\;,
\label{eqq1}
\end{equation}
\begin{equation}
\ddot \psi + 3\frac{\dot a}{a} \dot \psi - 12 \delta
\xi' \frac{\ddot a}{a} \left(\frac{\dot a^2}
{a^2}+\frac{1}{a^2}\right)+V'(\psi) =0\;,
\label{eqq2}
\end{equation}
with the
constraint
\begin{equation}
3 \left( \frac{\dot
a^2}{a^2}+\frac{1}{a^2}\right) (1+4\delta \dot \xi \frac{\dot
a}{a})=\frac{\dot \psi^2}{2}+V(\psi)\;.
\label{eqq3}
\end{equation}

    The dynamics  described by Eqs.(\ref{eqq1})-(\ref{eqq3})  is
more        complicated         in        comparison        with
Eqs.(\ref{constr})-(\ref{eqmot2}) due to the possibility for the
scale factor to have maxima  and  minima. We use the method  and
notations  of  \cite{Khalat7},  where  the  dynamics  of  closed
universe with the scalar field was studied in the absence of the
Gauss-Bonnet term.

    Two important regions of the configuration space $(a, \psi)$
can   be   easily   found   from   the   equations   of   motion
(\ref{eqq1})-(\ref{eqq3}): the region of possible extrema of the
scale  factor  $a$  (for  historical  reasons  it is called  the
Euclidean one) and the part of the Euclidean region where minima
of $a$ (i.e. the points of bounce) can appear.

\begin{figure}
\epsfxsize=0.6\hsize
\epsfysize=0.6\hsize
\centerline{\epsfbox{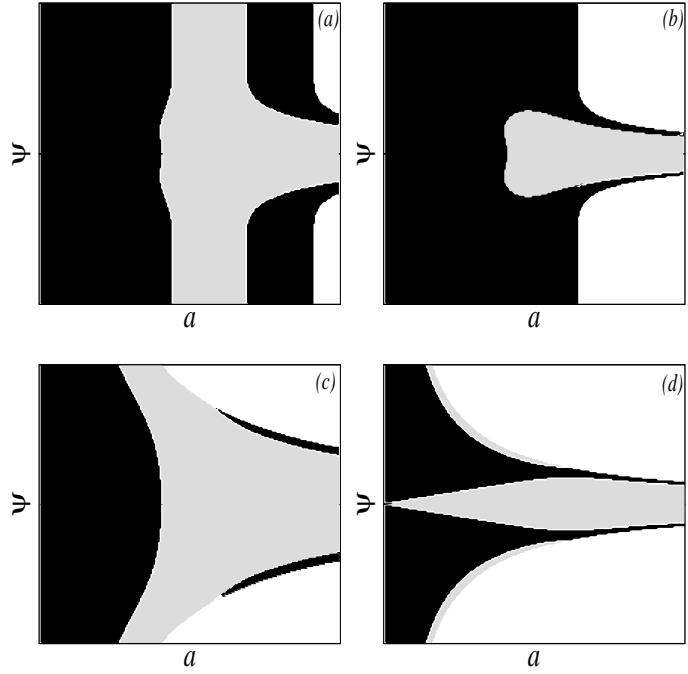}}
    \caption{Regions where  points  of  maximal expansion (gray)
and minimal  contraction  (black)  are  located in configuration
space  $(a,   \psi)$.   In   Fig.1(a)   $\xi=\psi^2$,   $V$   is
asymptotically flat  with  $\Lambda  <  \Lambda_0$;  in Fig.1(b)
$\xi=\psi^2$,  $V$   is  asymptotically  flat  with  $\Lambda  >
\Lambda_{0}$;  in  Fig.1(c)  $\xi=\psi^2$, $V=m^2 \psi^2/2$;  in
Fig.1(d)   $\xi=\psi^4$,   $V=m^2   \psi^2/2$.  All  plots   are
symmetrical with respect to transition $\psi \to -\psi$.}
\end{figure}

    Eq.(\ref{eqq3}) shows  that  the  Gauss-Bonnet term does not
change the configuration of the Euclidean region. From the other
hand, curves separating  the  points of maximal expansion ($\dot
a=0, \ddot  a <0$) and  the points of minimal contraction ($\dot
a=0,  \ddot   a  >0$)  change  significantly.  Substituting  the
condition $\dot  a=0$  into  (\ref{eqq1})  and  expressing $\dot
\psi$ from (\ref{eqq3}) we obtain the following equation for the
separating curve
\begin{equation}
-\frac{2}{a^2} + V(\psi) + \frac{4 \delta}{a^2} \xi'' (\psi)
\biggl(-\frac{3}{a^2} + V(\psi) \biggr)
+ \frac{2 \delta \xi' (\psi) V' (\psi)}{a^2} = 0\;.
\end {equation}

    We  start  with  the  case  $\xi=\psi^2$.  As  for  the flat
metrics,  this  case is exceptional one. Zero modulus  potential
leads to asymptotic (2) with  constant  scale factor $a = a_0  =
\sqrt{-12 \delta}$. The separating curve is simply the line $a =
a_0$.  All  the  terms  in  constraint  (\ref{eqq3})  tend  to a
constant  value  on this  asymptotic  and,  therefore,  only  an
asymptotically flat  potentials  could  not  destroy  it.  Small
enough $\Lambda$ corresponds  to  the situation of Fig.1(a), the
nonsingular  regime   is  preserved.  This  picture  changes  if
$\Lambda$  exceeds  the  value  $\Lambda_0 = (2  +  \sqrt{3})/(4
|\delta|)$  (Fig.1(b)).  In  this  case the regime $a  \to  a_0$
disappears  and  a  trajectory  which  describes  a  contracting
universe can either  fall into a  singularity or pass  trough  a
bounce and enter  an  expanding phase. Potentials unbounded from
above always violate the nonsingular asymptotic (2).

    Now we consider the important case  $V=m^2  \psi^2/2$  in  a
more detail (Fig.1(c)).

    The   equations    (\ref{eqq1})-(\ref{eqq3})   were   solved
numerically using the  method  of integration over an additional
parameter developed  in  Ref.  \cite{sts}. The maximal expansion
point was chosen  as  an initial  data  for the Cauchi  problem,
unknown  values  were  obtained  from  the  constraint  equation
(\ref{eqq3}).

    If we start  from the point  with $\dot a=0$,  initial  data
corresponding to  solutions  experiencing  at  least  one bounce
forms the system of quasiparallel zones in the initial condition
space ($a,\psi$). This system is rather similar to those studied
recently in the  closed FRW models without the Gauss-Bonnet term
\cite{Khalat7}, but has several important new features.

\begin{figure}
\epsfxsize=0.6\hsize
\epsfysize=0.8\hsize
\centerline{\epsfbox{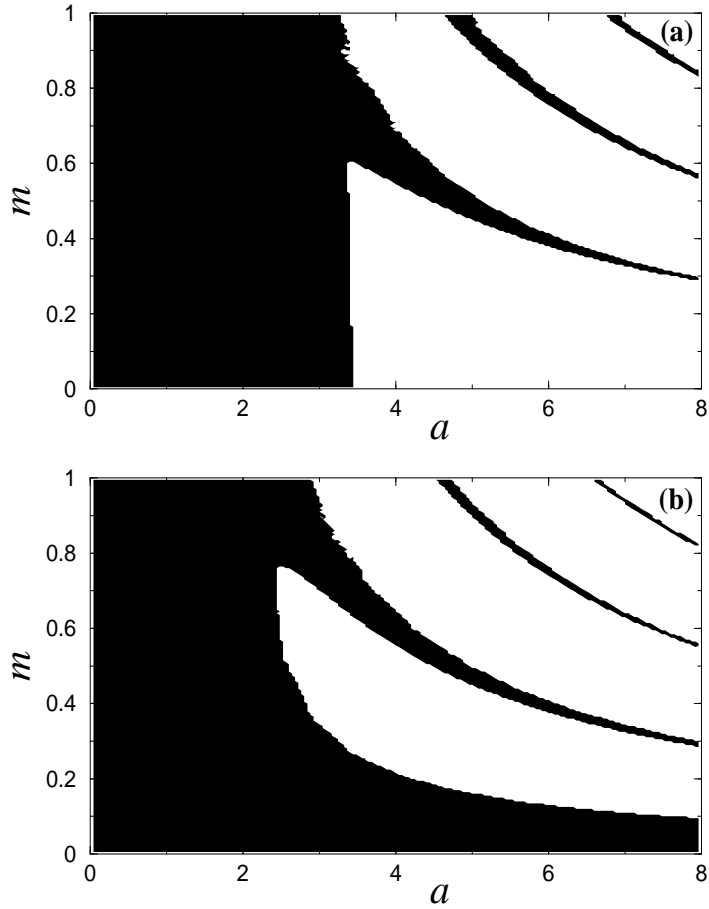}}
    \caption{The  $\psi=0$  cross-sections  of zones in  initial
condition  space  $(a,  \psi)$  leading  to  bounce  for  $V=m^2
\psi^2/2$.    In     Fig.2(a)    $\xi=\psi^2$,    in     Fig2(b)
$\xi=cosh(\psi)$}
\end{figure}

    First  of  all,  it  is  clear  from  the  separating  curve
configuration  that  if the initial scale factor  is  less  than
$\sqrt{-12 \delta}$,  the  trajectory  starts  from  the  bounce
itself. So, the  1-st bounce interval  is adjacent to  the  axis
$a=0$ in contrast to the results  of Ref.\cite{Khalat7}. Another
important  feature  can be seen from the  Fig.2(a).  We  plotted
cross-sections $\psi=0$ of the bounce intervals depending on the
mass $m$. If  $m<0.6$  we  can see the full  set  of  intervals:
trajectories   from   1-st  interval  have  a  bounce  with   no
$\psi$-turns before it, trajectories which have initial point of
maximal expansion between 1-st and  2-nd  intervals  fall into a
singularity after one $\psi$-turn, those from 2-nd interval have
a bounce after 1 $\psi$-turn  and  so on. If $m$ becomes  bigger
than  $0.6$,  the  2-nd  interval contains trajectories  with  2
$\psi$-turns before  bounce,  the  space  between  1-st interval
(which is now the product of two merged intervals) and  the 2-nd
one contains trajectories  falling  into a singularity after two
$\psi$-turns. There are  no  trajectories going to a singularity
with exactly one $\psi$-turn.

    This process of interval merging continues  with the modulus
field mass  $m$ increasing. The picture  is similar to  the case
without the Gauss-Bonnet term but with  gently sloping potential
\cite{serg},  where  the process of interval merging also  takes
place. In both cases, the  interval  formed  by merging contains
very chaotic  trajectories which can not  be described in  a way
similar to  \cite{Khalat7}.  A  significant part of trajectories
starting from such an interval  may  oscillate  during very long
time without falling into a singularity or leaving the interval.
In Fig.3. we present the number of trajectories  having at least
50 oscillations of  modulus field versus the modulus field mass.
The initial scale factor varies with step $0.02$ in the range of
scale  factors  from  $1$  to   $5$   units   (total  number  of
trajectories for each value of modulus mass is  equal to $200$).
We ignore trajectories which describe a long-time expansion (the
calculations stopped  when  running  scale  factor  was 10 times
greater than  the  initial  one),  so,  all counted trajectories
represented the chaotic oscillations. A special  effort was done
\cite{Corn-Shel}  to   construct very  chaotic  solutions  in  a
closed FRW model without the Gauss-Bonnet term, and trajectories
having at most $12$ oscillations of scalar field  were found. We
choose a much bigger value ($50$ oscillations), so the existence
of  such  trajectories  tells  us  that   a  qualitatively  more
complicated      chaos      exists       in      the      system
(\ref{eqq1})-(\ref{eqq3}).  If  $m<0.7$  the  measure  of   such
trajectories is  so small, that they do not  appear on our grid.
For bigger  $m$ this measure  grows rather rapidly. For the mass
exceeding the first merging  value  $1.3$ times more than $30\%$
of  trajectories  having  the  point  of  maximal  expansion  in
$\psi=0$ cross-section of  the  first interval are very chaotic.
So, the regime of  chaotic  oscillation for modulus field masses
exceeding the first  merging value is much more significant than
for the case without the Gauss-Bonnet term.

    This picture does not  change  qualitatively if we switch on
the  self-interaction  of  the  modulus  field.   For  the  pure
self-interacting   potential    $V(\psi)=\lambda   \psi^4$   the
corresponding critical value  of  $\lambda$ leading to the first
bounce interval merging is equal to $\sim 1.2$.

\begin{figure}
\epsfxsize=0.6\hsize
\epsfysize=0.6\hsize
\centerline{\epsfbox{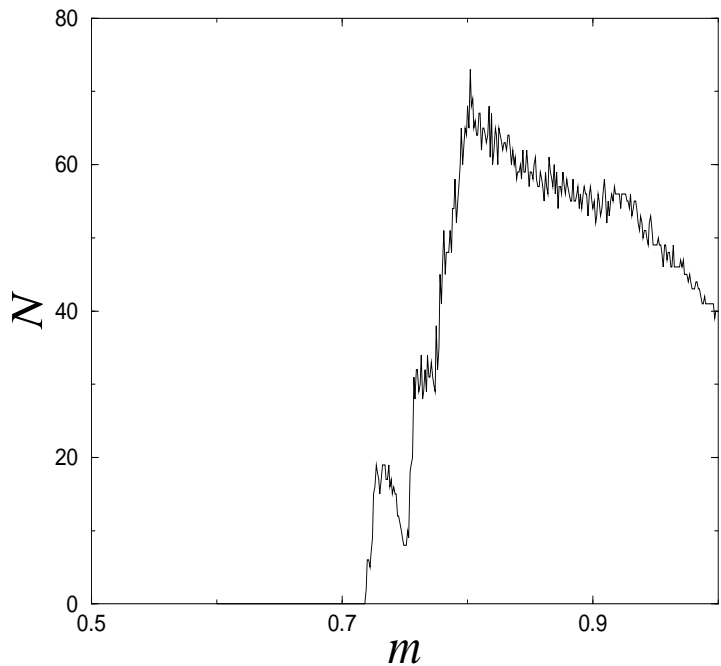}}
    \caption{The number of trajectories  avoiding  a singularity
during first 50  oscillations of the modulus field. Total number
of  trajectories  is  equal  to 200, initial scale  factors  are
located in the range of two first bounce intervals}
\end{figure}

    When  $\xi(\psi)$  grows faster than $\psi^2$, there are  no
solution with asymptotically constant scale factor even with the
vanishing modulus field potential. For $\xi \sim \psi^{n+1}$ the
asymptotic solution is
\begin{eqnarray}
a(t) & = & \sqrt{\frac{n+3}{2n}}(n+1)^{\frac{n-1}{n+1}} (-4n\delta)
^{\frac{1}{n+1}}
t^{\frac{n-1}{n+1}}\;, \nonumber \\
\psi(t) & = & \left(-\frac{(n+1)^2}{4n\delta}\right)^{\frac{1}{n+1}}
t^{\frac{2}{n+1}}\;, \label{eqq4}
\end{eqnarray}
    for $\xi \sim  exp(|\psi|)$ the solution was found by Easter
and Maeda in \cite{m3}. In both cases it  describes the universe
with growing scalar  factor  and modulus field. The Gauss-Bonnet
term tends to  zero  and, therefore,  at  some point it  becomes
unimportant in comparison with  arbitrary  non-vanishing modulus
field potential. After that, the evolution of the universe obeys
the well-known rules for FRW  dynamics  with  the massive scalar
field which inevitably lead to recollapse.

    Now it  is time to outline  the bouncing properties  in this
case. The separating curve  looks  like plotted in Fig.1(d). But
the $\psi=0$ cross-section of bouncing intervals  is not changed
much in comparison with the $\xi \sim \psi^2$ case. Again, there
is the first interval in the range of very small  scale factors,
which is now becoming thinner  with  the  growing modulus field.
The merging of the bouncing  intervals  leading  to  significant
chaotisation of trajectories is also takes place (Fig.2(b)).

    Additionally we  conclude  from  Fig.2(b)  that  though  the
asymptotic (\ref{eqq4})  for  zero modulus potential is regular,
taking into  account  the  classical  evolution  of the Universe
trough the maximal expansion  point  back to the large curvature
regime makes the  dynamics chaotic with the strong dependence on
the initial condition.  Similar to the $\xi=\psi^2$ case we have
no regular nonsingular solutions, but there  are infinite number
of unstable chaotic  trajectories  escaping a singularity for an
arbitrary long time.

\section{Conclusions}

    In this paper  we  investigated the cosmological behavior of
the model with the  second  order curvature corrections based on
the  action  (\ref{a1})  for  flat  and  closed  FRW  metrics. A
particular attention was paid to the  possibility of nonsingular
past     asymptotics.      We     confirm     recent      claims
\cite{Antoniadis:1994jc,Rizos:1994rt,Kanti:1998jd}   that   such
regimes exist at least for functions $\xi(\psi)$ with $\xi'' >0$
and   present   corresponding   formulas   for   power-law   and
exponentially steep  $\xi(\psi)$. The previously known case $\xi
\sim \psi^2$ appears to be in some sense the exceptional one. It
corresponds  to  DeSitter flat universe or closed universe  with
the constant  scale  factor  (see Eqs.(2),(\ref{psi2})). Steeper
$\xi(\psi)$ lead to power-low dependence of the Hubble parameter
upon  time  for  the  flat  metric  and  to  bounce  nonsingular
solutions  for  the  closed  one  (a   particular  example  with
exponentially steep $\xi(\psi)$ for the  latter  case  was  also
studied by Easther  and Maeda in \cite{m3}). The important point
also mentioned in above cited  papers  is  that these asymptotic
regimes do not require a fine-tuning of initial conditions.

    This optimistic  view  on  the  nonsingular regimes changes,
however, if we take into account  possible  potential  term  for
modulus field, which naturally arises in the description of SUSY
breaking. We  conclude  that  a  modulus  potential destroys the
nonsingular asymptotics in most cases we  have  studied.  For  a
flat  universe  the  only  case  with  $\xi''  >0$  which allows
nonsingular  regime  and a massive modulus field  is  $\xi  \sim
\psi^2$. Even in this case a steeper modulus potential kills the
nonsingular asymptotic. For $\xi \sim  \psi^4$  it  can  survive
only for asymptotically  flat  potentials. Steeper $\xi$ can not
prevent  from  falling into  a  singularity  with  an  arbitrary
modulus  field  potential not tending to zero  for  $|\psi|  \to
\infty$.  Our   numerical   simulations   show   that  when  the
nonsingular regime  described  above becomes impossible, all the
cosmological  trajectories  go towards a singularity and no  new
nonsingular regime appear.

    For a closed universe the possibility of regular nonsingular
asymptotic  is   even   more  restricted.  The  above  mentioned
asymptotic can survive only in  the  exceptional  case $\xi \sim
\psi^2$  and  an  asymptotically  flat  potential.   But  now  a
trajectory which does not belong to this regime may  either fall
into  a  singularity or  experience  bounce,  depending  on  the
initial conditions.  At  the  expanding  phase  after bounce the
Gauss-Bonnet term become irrelevant  and  we can use the results
obtained for the scalar field in Einstein theory.  Since for the
closed  metric  all the trajectories must have  their  point  of
maximal expansion, a bouncing universe will begin to contract at
some point  and, at the second time, fall  into a singularity or
have a  second bounce etc. So, we return  to the chaotic picture
similar to one described for  the  massive  scalar field coupled
with the Einstein gravity. This chaotic regime is very sensitive
to  the  initial  condition  and  does  not  contain any  stable
nonsingular trajectory.

    For  steeper   $\xi(\psi)$  the  regime  without  a  modulus
potential  is  a bounce itself, so arbitrary  potential  at  the
after-bounce stage will dominate the Gauss-Bonnet  term and lead
to recollapse and chaotisation of trajectories.

    Since  the   particular   form  of  second  order  curvature
corrections arising from the string theory include exponentially
steep $\xi(\psi)$  with  coming from SUSY breaking exponentially
steep modulus potential, nonsingular solutions presenting in the
string gravity with $V(\psi)=0$ disappear when modulus potential
is taken into account.

\section*{Acknowlegements}
    This work was partially supported by  Russian Foundation for
Basic  Research  via grant N 99-02-16224. A.T.  is  grateful  to
T.Damour for useful discussion.


\section*{References}

\begin{thebibliography}{99}

\bibitem{s2}
M.B.Green, J.H. Schwartz and E.Witten,
``Superstring Theory'',
Cambridge University Press, Cambridge (1986).

\bibitem{s1}
J.H. Schwarz, {\it Phys. Rep.} {\bf 315}, 107 (1999).

\bibitem{m1}
S. Mignemi and N.R. Stewart, {\it Phys. Rev.}
{\bf D 47}, 5259 (1993);

\bibitem{k1}
P. Kanti, N.E. Mavromatos, J. Rizos, K. Tamvakis and E. Winstanley,
{\it Phys. Rev.} {\bf D 54}, 5049 (1996);

P. Kanti and K. Tamvakis,
{\it Phys. Lett.} {\bf B 392}, 30 (1997).

\bibitem{m3}
R. Easter and K. Maeda, {\it Phys. Rev.} {\bf D 54}, 7252 (1996).

\bibitem{sts}
S.O. Alexeyev and M.V. Pomazanov,
{\it Phys. Rev.} {\bf D 55}, 2110 (1997);

S.O. Alexeyev and M.V. Sazhin,
{\it Gen. Relativ. and Grav.} {\bf 8}, 1187-1203 (1998).

\bibitem{m2}
T. Torii, H. Yajima and K. Maeda,
{\it Phys. Rev.} {\bf D 55}, 739 (1997);

R.C. Myers, ``Black Holes in Higher Curvature Gravity'',
in {\it ``Black Holes, Gravitational Radiation and the Universe:
Essays in Honor of C.V. Vishveshwara''}, eds C.V. Vishveshwara,
B.R.Iyer, B.Brawal, gr-qc/9811042.

\bibitem{x1}
T. Banks, Nucl. {\it Phys. Proc. Suppl.} {\bf 67}, 180 (1998);

J.H. Schwarz and N.Seiberg, {\it Rev. Mod. Phys.}
{\bf 71}, S112 (1999).

\bibitem{tj1}
T.Jacobson, ``Black Hole Thermodynamics Today'',
    {\it Talk, given at 8th  Marcel  Grossman  Meeting on Recent
Developments in Theoretical and Experimental General Relativity,
Gravitation and Relativistic Theories (MG8), Jerusalem,  Israel,
22-27 June 1997},
gr-qc/9801015.

\bibitem{Antoniadis:1994jc}
I.~Antoniadis, J.~Rizos and K.~Tamvakis,
{\it Nucl. Phys.} {\bf B415} 497 (1994).

\bibitem{Rizos:1994rt}
J.~Rizos and K.~Tamvakis,
{\it Phys. Lett.} {\bf B326}, 57 (1994).

\bibitem{Kanti:1998jd}
P.~Kanti, J.~Rizos and K.~Tamvakis,
{\it Phys. Rev.} {\bf D59}, 083512 (1999).

\bibitem{Khalat7}
A.Yu.Kamenshchik, I.M.Khalatnikov, A.V.Toporensky,
{\it Int. J. Mod. Phys.} {\bf D6}, 673 (1997).

\bibitem{Tseytlin}
A.Tseytlin,
``String  Solutions  with  Nonconstant Scalar Fields''
    {\it Published in the proceedings of International Symposium
on Particle  Theory,  Wendisch-Rietz,  Germany,  7-11  Sep  1993
(Ahrenshoop Symp.1993:0001-13),}
hep-th/9402082

\bibitem{serg}
S.A.Pavluchenko and A.V.Toporensky,
``Chaos in FRW Cosmology with Gently
Sloping Scalar Field Potentials'',
gr-qc/9911039.

\bibitem{Foster}
S.~Foster,
``Scalar Field Cosmological Models With Hard Potential Walls,''
gr-qc/9806113.

\bibitem{Corn-Shel}
N.J. Cornish, E.P.S. Shellard,
{\it Phys. Rev. Lett.} {\bf 81}, 3571 (1998).

\end{thebibliography}
\end{document}